# Molecular dynamics of ice-nanotube formation inside carbon nanotubes

Junichiro Shiomi[1], Tatsuto Kimura[2] and Shigeo Maruyama[1,*]

[1]Department of Mechanical Engineering, The University of Tokyo
7-3-1 Hongo, Bunkyo-ku, Tokyo 113-8656, Japan

[2]Department of Mechanical Engineering, Kanagawa University,
3-27-1 Kanagawa-ku, Yokohama-shi, Kanagawa, 221-8686, Japan

The first order phase transition of a water cluster confined in a dynamic single-walled carbon nanotube is investigated using a classical molecular dynamics (MD) method. The formation of ice-nanotube is monitored through the structure factor and potential energies. The transition temperature and its diameter dependence obtained by the simulations agree well with those of previously reported experiments. The transition temperature of the ice-nanotube was shown to take a maximum value of around room temperature with the number of the ring members $n$=5. Potential energy contribution to the phase change is generally dominated by that of the intrinsic water-water interaction, while that of water-carbon interaction plays a significant role on determining the dependence of transition temperature on the nanotube diameter.

*Corresponding author: Tel: +81-3-5841-6421, Tel: +81-3-5800-6983.
E-mail address: maruyama@photon.t.u-tokyo.ac.jp

## I.   INTRODUCTION

Investigation of water confined in low dimension holds much importance as it is a key system in bioscience[1] and nanotechnology under aqueous environments. The confinement is expected to alter the phase transition and various transport properties of water from those of bulk water. An ultimate realistic case of such low-dimensional systems is water confined in single-walled carbon nanotubes (SWNTs). An anomalous phase transition of water to ice-nanotube was first predicted by classical molecular dynamics (MD) simulations of water inside a static SWNT under high pressure (axial pressure of 50-500 MPa)[2]. It was found that the water experiences a first order phase transition to form an ice-nanotube (I-NT), where the number of members of the circumferential ring ($n$) was determined by the diameter ($d$) of the surrounding SWNT. Molecular simulations have been also used to explore detail structures of the confined water and their energetic properties[3,4]. The existence of ice-nanotubes was first confirmed in experiments using the X-ray diffraction analyses[5,6]. Experiments were performed at around the saturated vapor pressure, where condensation of water inside SWNTs occurred around 315-330 K. The experiments delivered a striking feature of the phase change that the ordering transition temperature increases, even to room temperature, as the nanotube diameter i.e. $n$ of I-NT decreases. This trend is opposite from that of the bulk water in a glass capillary tube[7] and hence indicates a crossover of physics from bulk to atomic scale phenomena on reducing the diameter[6].

In the current study, we investigate the diameter dependence of the transition temperature of a saturated water cluster locally confined in an SWNT by monitoring instantaneous molecular structures and potential energy. While the transition temperature has been calculated for various nanotube diameters (or $n$) by MD simulations[2] and Grand canonical Monte Carlo simulations[4], the current work is first to provide the direct comparison with the experiment in terms of the dependence of the transitional temperature on the nanotube diameter by simulating the realistic phase-change process without artificial pressure treatments. Unlike the earlier MD calculations of the transition[2], the model includes the carbon-carbon interaction dynamics based on a potential function that have been shown to exhibit the phonon density of states of carbon nanotubes with a sufficient accuracy[8-10].

## II. MOLECULAR DYNAMICS SIMULATIONS

Water molecules were expressed by SPC/E potential[11]. SPC/E potential is expressed as the superposition of Lennard-Jones function of oxygen-oxygen interaction and the electrostatic potential by charges on oxygen and hydrogen as follows,

$$\phi_{12} = 4\varepsilon_{OO}\left[\left(\frac{\sigma_{OO}}{R_{12}}\right)^{12} - \left(\frac{\sigma_{OO}}{R_{12}}\right)^{6}\right] + \sum_i \sum_j \frac{q_i q_j e^2}{4\pi\varepsilon_0 r_{ij}}, \quad (1)$$

where $R_{12}$ represents the distance of oxygen atoms, and $\sigma_{OO}$ and $\varepsilon_{OO}$ are Lennard-Jones parameters. The Coulombic interaction is the sum of 9 pairs of point charges. The SPC/E potential is known to predict correct phase change temperature and is widely used in micro/nano heat transfer[12]. The potential function between water molecules and carbon atoms were represented by Lennard-Jones function of the distance between the oxygen in the water molecule and the carbon atom. Here, we ignored the quadropole interactions as they were found to have minute impact[13].

The carbon interactions were expressed by the Brenner potential function[14] in a simplified form[15], where the total potential energy of the system is modeled as,

$$E = \sum_i \sum_{j(i<j)} \left[V_R(r_{ij}) - B^*_{ij} V_A(r_{ij})\right]. \quad (2)$$

Here, $V_R(r)$ and $V_A(r)$ are repulsive and attractive force terms, which take the Morse type form with a certain cut-off function. $B^*_{ij}$ represents the effect of the bonding condition of the atoms. As for the potential parameters, we employ the set that was shown to reproduce the force constant better (table 2 in Ref. 14). It has been shown that the formulation of potential function exhibits phonon dispersion with sufficient accuracy[8-10]. The inclusion of the lattice vibrations of carbon nanotubes enables us to incorporate the realistic heat transport from an SWNT to the water cluster. The thermal boundary conductance between the SWNT and the confined liquid water was previously computed to be typically about 5 MW/m$^2$K[16].

A typical simulation begins with an initial condition with liquid water locally confined in an SWNT. Figure 1(a) shows the picture of a water cluster of 192 molecules adsorbed inside a (9, 9)-SWNT at room temperature or higher. The system was subjected to a periodic boundary condition in the axial direction. The cast SWNT ($L$=20.2 nm) is longer than sum of the cut-off distance of the Coulombic potential and the length of water cluster and hence there is no inter-cluster interaction through the periodic boundary. The cut-off distance of the Coulombic potential was set to be 10 nm, except for the case of (8, 8) SWNT, where the water-cluster length exceeds half the nanotube length. In case of (8, 8) SWNT, in order to avoid the inter-cluster interaction, the cut-off distance of the Coulombic potential was reduced to 5 nm. For this case, we have also preformed a simulation for cut-off distance of 2.5 nm and confirmed that the phase transition characteristics exhibit negligible dependence on the cut-off distance in the current range. The velocity

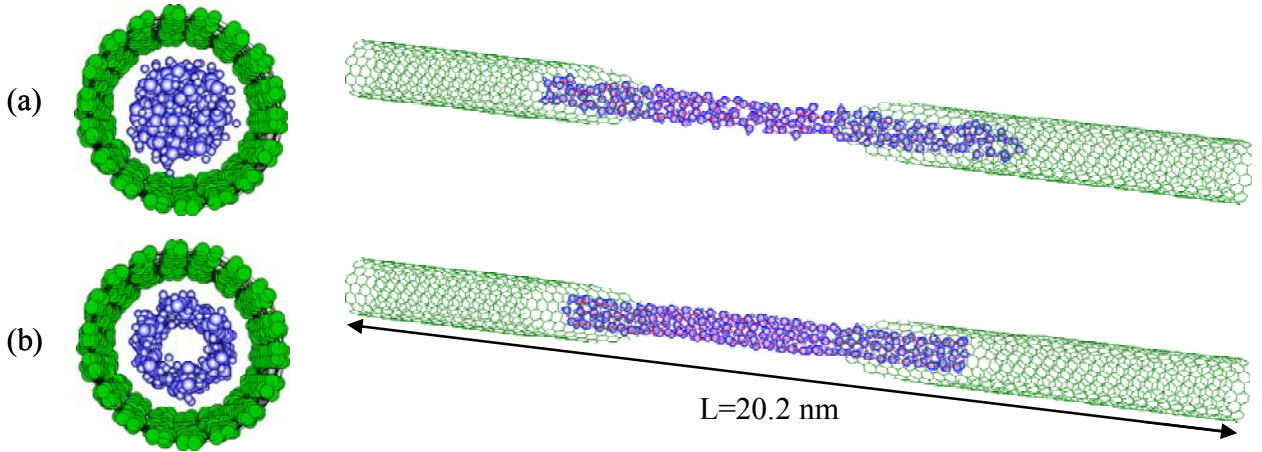

Figure 1. (a) a typical initial condition of a saturated water cluster in a (9, 9) SWNT and (b) the ice-nanotube crystallized by cooing the SWNT to 200 K ($n$=6).

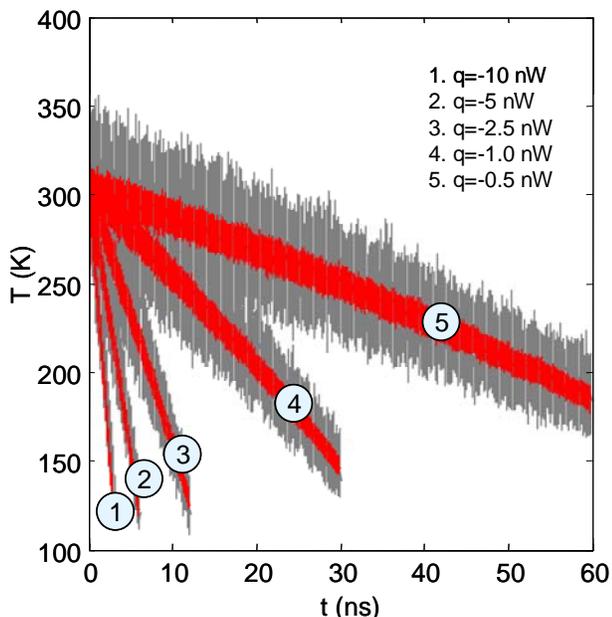

Figure 2. Typical time histories of spatially averaged temperatures of an SWNT (solid line) and the confined water cluster (dotted line) for various values of the heat flux $q$.

Verlet method was adopted to integrate the equation of motion with the time step of 0.5 fs. After initially equilibrating the system at around 300 K, the temperature of the system was gradually decreased by applying isotropic constant heat flux to the SWNT. With negative heat flux $q$, heat was subtracted as

$$v_i = v_i \sqrt{1 + q/E_k} \qquad (3)$$

where $E_k$ is the total kinetic energy of the SWNT. Figure 2 shows time histories of the resulting temperature of a (9, 9)-SWNT and the confined water cluster for various values of $q$. Once the crystal is formed, a melting numerical experiment was performed with positive $q$ in Eq. (3) to examine the hysteresis of the transition.

As the casts to produce ice-nanotubes, we use armchair SWNTs with chiral-index pairs of (7, 7), (8, 8), (9, 9) and (10, 10) with diameters of $d$=0.97, 1.11, 1.25 and 1.39 nm, which correspond to I-NTs with $n$=4, 5, 6 and 8, respectively. To fill in the absence of an I-NT with $n$=7, an additional case was carried out for (10, 9)-SWNT ($d$=1.32 nm). A cluster consists of 192 water molecules except for (7, 7) SWNT, where the number of water molecules was reduce to 96 due to the enhanced computational cost attributed to the relatively long elongation of the cluster in liquid phase. Although simulations were performed for values of d smaller than 0.96 nm, we did not observe formation of ice-nanotubes with $n \leq 3$ in the temperature range T≥100 K.

## III. RESULTS AND DISCUSSIONS

The crystallization of water can be directly monitored in the radial density distribution of water molecules. Figure 3(a) exhibits radial density distribution functions of oxygen atoms $g_{oo}(r)$ averaged over certain time durations in the case of $n$=6. Radial density profiles are shown for water at (I) pre-transition, (II) transition and (III) post-transition stages to demonstrate the crystallization process on cooling the SWNT with $q$=-1 nW. The appearance of the local peaks corresponds to the interatomic length of the crystal. In addition to the equidistant peaks originated from the axial alignment of the oxygen atoms, there are distinct non-equidistant peaks below $r$=10 Å due to the circumferential structure of the ice-nanotube. By performing Fourier transform, we obtain the structural index $S(k)$ which allows us to carry out a direct comparison with the X-ray diffraction experiments[5,6]. The spectra $S(k)$ computed from $g_{oo}(r)$ profiles in Fig. 3(a) are shown in Fig. 3(b). The peak corresponding to the axial bond-length of the ice-nanotube, though not shown here for brevity, shifts slightly depending on $n$ reflecting the structural variation of I-NTs. The peak intensity of $S(k)$ indicates the ordering process of the I-NT as shown in Fig. 3(c). The pictures visualize the transition process where the crystal locally nucleates and grows in time. Figure 3(d) shows the time history of the peak intensity, which was quantified by integrating the local spectrum $S(k)$ around the peak. The intensity nearly saturates after the completion of the crystallization, though it grows slightly even below the freezing temperature due to the reduction of the thermal lattice vibration.

The first order liquid-solid phase transition can also be detected by monitoring the potential energy profiles[3,4]. The calculation of the scalar instantaneous potential energy is simpler and takes much smaller memory compared with the above-mentioned calculation of the structural factor with sufficient resolution in wave number. Figure 4(a-c) shows the potential energy per number of water molecules computed from (a) the water-water intrinsic potential ($V_{ww}$), (b) water-carbon interfacial interaction potential ($V_{wc}$) and (c) the total potential energy of the water ($V_t = V_{ww} + V_{wc}$). In all the figures, the phase change can be detected in the jump $\Delta V=V_{liquid}(T_m+\Delta T_m/2)-V_{solid}(T_m-\Delta T_m/2)$, where $T_m$ is the transition (melting) temperature and $\Delta T_m$ is the extent of hysteresis. Here, $T_m-\Delta T_m/2$ and $T_m+\Delta T_m/2$ are the temperatures at which the freezing and melting transitions are completed in the cooling and heating numerical experiments, respectively. Comparison of magnitudes of $\Delta V_{ww}$ and $\Delta V_{wc}$ indicates that the phase change is encouraged dominantly by the energy gain in the water-water interaction. Note that the magnitude of the interaction potential energy of the water and the hydrophobic SWNT ($V_{wc}$) is one order smaller than that of the intrinsic water-water interaction ($V_{ww}$).

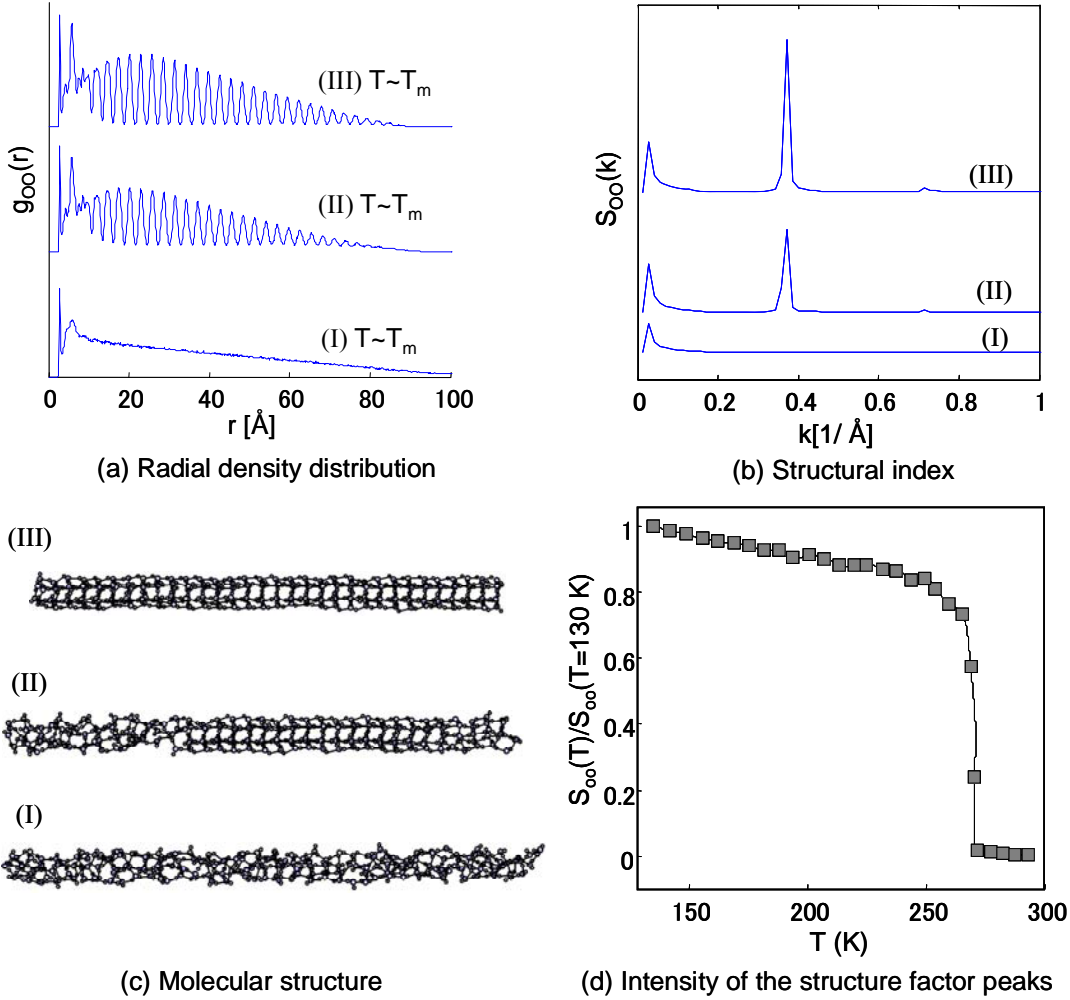

Figure 3. (a) Radial density distributions of oxygen atoms of a water cluster confined in a (9, 9) SWNT at different stages of crystallization; (I) $T>T_m$, (II) $T\sim T_m$ and (III) $T<T_m$. (b) The corresponding structure factor $S(k)$. (c) Visual images of the molecular structures. (d) The intensity of $S(k)$ (normalized with $S(k)$ at $T$=130 K ) at different temperature.

In Fig. 4(a-c), for each value of $d$, two lines of cooling and heating numerical experiments are drawn and marked with arrows to highlight the extent of the hysteresis, $\Delta T_m$. In the MD simulations, the restriction on the computational time limits the magnitude of heat flux $|q|$ to be much larger than reality. Therefore, the water cluster can be easily under-cooled in the cooling process due to the comparable time scales between the cooling of the system and the crystal growth of the ice-nanotube, and vice versa for the heating process. For a certain value of $n$, $\Delta T_m$ decreases with $|q|$ and should eventually diminish when $|q|$ is small enough. As shown in Fig. 4, with the smallest $|q|$ (=0.5 nW) in the current study, $\Delta T_m$ could be reduced to some extent, though the hysteresis could not be completely eliminated. As will be discussed later, in case of a (10, 10) SWNT, I-NT was formed only under certain simulation conditions. The profiles for I-NT with $n$=8 presented in Fig. 4(a-c) are obtained at the cooling speed of 5 K/ns.

In Fig. 5, the $d$-dependence of the transition (melting) temperature $T_m$ calculated based on the above potential energy analyses is shown together with the experimental results[6]. The data are accompanied with sketches of the cross-sections of the corresponding I-NTs. Let us now focus on $n \geq 5$, where experimental results are available. The simulation reproduces a key quantitative feature observed in the experiments[6], where water freezes at around room temperature in case of $n$=5. The agreement becomes worse for larger $n$, which may have to do with the increasing influence of rigid SPC/E water model as the I-NT becomes unstable. The agreement is rather surprising on considering the expected size effect of the stability. Koga et al.[3] showed for static I-NTs ($T$=0 K) expressed with CC and TIP4P potential models that the energy per molecule decreases by 3-6 kJ mol$^{-1}$, depending on $n$, on varying $N$ from 15 to infinity. Since the size-dependence of potential energy may influence $T_m$, careful investigation of the size effect would be necessary before concluding the quantitative agreement with the experiments. However, a thorough exploration of the size effect with the current approach including the interfacial dynamics would require considerable computational cost therefore remains as our future task.

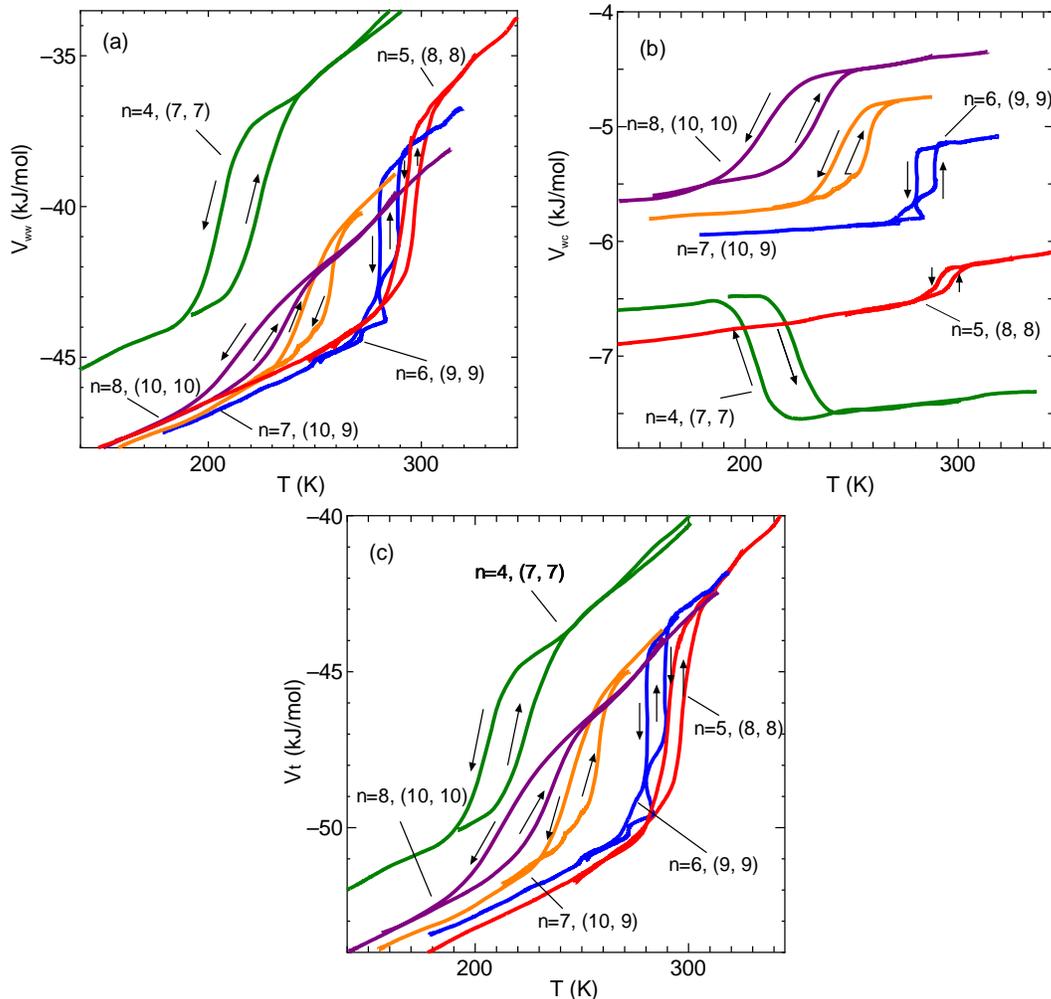

Figure 4. Temperature dependences of (a) water-water ($V_{ww}$), (b) water-carbon ($V_{wc}$) and (c) total ($V_t$) potential energies for various SWNT-diameters ($d$) and numbers of ring-members ($n$). Hysteresis is highlighted by plotting the profiles of both cooling and heating processes as denoted by arrows.

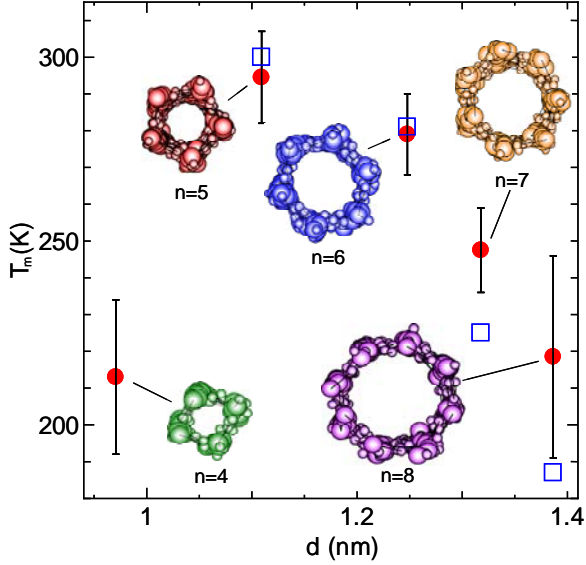

Figure 5. The transition (melting) temperature $T_m$ for various numbers of ring-members of ice-nanotubes ($n$). Simulation results (circles) are compared with the experimental results (squares)[6]. The error bars indicate the extent of hysteresis $\Delta T_m$.

The simulation successfully reproduces the qualitative trend observed in the experiment i.e. $T_m$ increases as $d$ decreases[6]. As discussed in Ref. 6, the trend is in contrast to the capillary freezing point depression of water confined in glass microcapillaries, where $T_m$ scales with $1/d$. In a three-dimensional capillary, $T_m$ is determined by the balance of the volume free energy and interfacial free energy whose dimension difference comes out as the scale $1/d$ (Ref. 7). The scaling apparently does not hold when the tube is so thin that the confined water cluster consists mostly of surface. Furthermore, unlike the three-dimensional bulk ice, ice under quasi-one-dimensional confinement takes variety of stable crystal structures depending on $d$ and hence the variation of the structural stability of I-NT is expected to influence the transition[6]. The current observation supports the idea, where the $d$-dependence of $V_t$ in solid phase [Fig. 4(c)] is in correlation with that of $T_m$ (Fig. 5) in $n \geq 5$. On the other hand, $V_t$ of liquid water shows minor dependence on $d$. It is worth pointing out that, as seen in Fig. 4(a), the intrinsic potential of water-water interaction[4] does not solely explain the above correlation. It can be seen in Fig. 4(b) that $d$-dependence of interfacial potential $V_{wc}$ needs to be considered to achieve the correlation.

In the current simulations, an I-NT with $n=4$ was observed, which gives rise to a maximum value of $T_m$ at $n=5$. I-NT with $n \leq 4$ has not been found in the experiments[6]. This could be due to the limited range of SWNT diameters in the experiment where the nanotubes of diameter from 1.09 to 1.52 nm were selectively chosen. It may also be due to the relatively unstable liquid water inside a (7, 7) SWNT. As shown in Fig. 4(c), the water inside a (7, 7) SWNT is less stable than the rest of the cases. This agrees with Grand canonical Monte Carlo simulations[17] where the critical $d$ of a pore sieving effect for SPC/E water molecules adsorbed inside an SWNT was observed to lie in between $d=0.81$ nm and 1.08. As seen in Fig. 4 (a-b), the fractions of contribution from $V_{ww}$ and $V_{wc}$ on the total potential energy shows that the relatively unstable feature of water inside a (7, 7) SWNT resides in the water-water interaction potential.

In the course of running numerical simulations, structure selection of system with (10,10)-SWNT was found to be sensitive to slight changes in simulation conditions such as cooling speed and temperature control method. Depending on the conditions, two different ice structures were obtained as illustrated in Fig. 6; an I-NT [Fig. 6 (I)] and an ice-nanotube with the hollow filled with water molecules lined along the axis [Fig. 6 (II)]. The latter structure has been discussed in connection with the unusually volatile dynamics of the confined water[18]. On monitoring $V_t$ during the formations of the two ice structures in (10, 10) SWNT [Fig. 7(a)], I-NT (I) was found to be more stable than the other ice structure (II). The energetic properties ($V_{ww}$ and $V_{wc}$) show that most of the stability difference is attributed to $V_{wc}$, the interfacial potential energy. The above observation of sensitivity on the simulation condition may suggest that the stability difference be counteracted by the relatively large entropy of structure (II) and consequently the free energy of the two structures become similar.

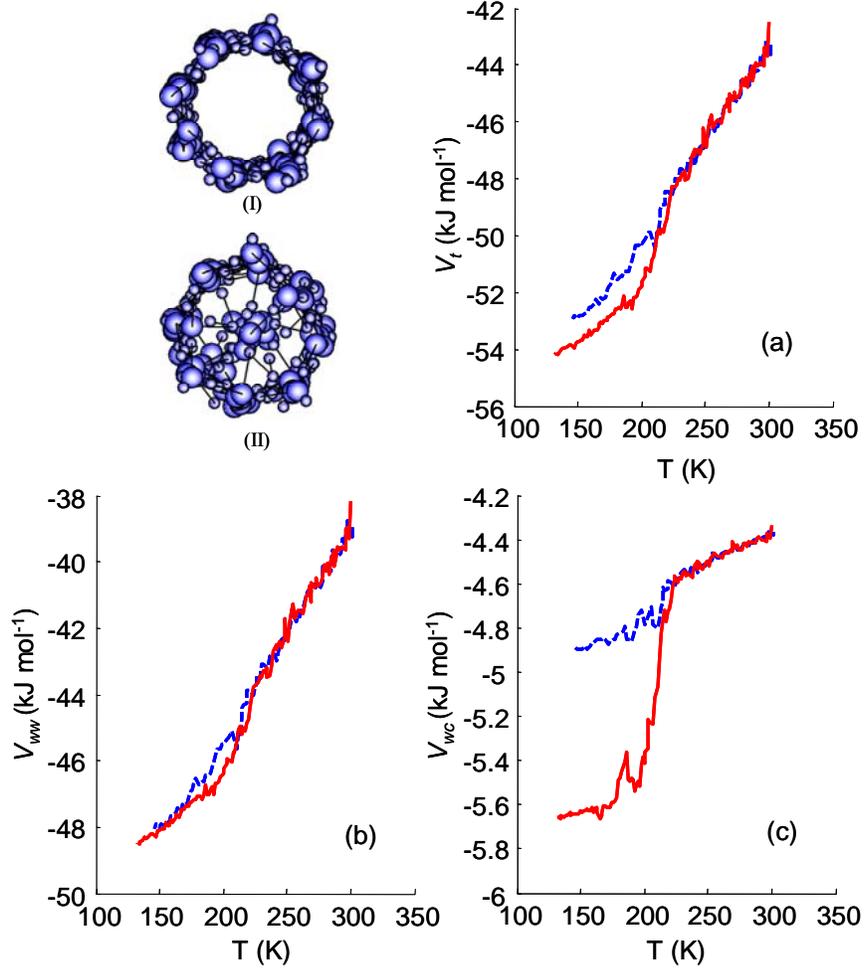

Figure 6. Stability comparison of two water-cluster structures (I) and (II). Figures show (a) the total potential energy ($V_t$), (b) the contribution from water-water interactions ($V_{ww}$) and (c) that from water-carbon interactions ($V_{wc}$). Solid and dashed lines denote the profiles of structures (I) and (II), respectively.

## IV. CONCLUSIONS

The first order phase transition of a water cluster confined in an SWNT to an ice-nanotube was investigated using a classical molecular dynamics method. By performing non-equilibrium simulations with constant heat flux, the phase transition was identified by monitoring the structure factor and potential energies. Simulations were preformed for various numbers of ring members $n$ (4-8) and transition (melting) temperature exhibited a maximum value at $n=5$. For $n=5$, $T_m$ was confirmed to reach as high as around room temperature. For $n \geq 5$, the melting temperature obtained from the simulations agrees well with that of the experiments not only qualitatively but also quantitatively despite the difference in the size of the system. The simulation confirms the limit of the ice-nanotube formation to be (10, 10) SWNT, beyond which ice forms a tube with the core filled with a chain of water molecules. The potential energy contribution to the phase change is generally dominated by that of water-water interactions. However, interfacial (water-carbon) potential energy plays significant role on determining the dependence of $T_m$ on $d$ ($n$) and relative stability of the tubal and non-tubal ice structures in (10, 10) SWNT, where the two structures exhibit similar intrinsic stability.

## ACKNOWLEDGEMENT

This work is supported in part by the Japan Society for the Promotion of Science for Young Scientists #1610109 and Grants-in-Aid for Scientific Research #17656072.

## REFERENCES
1. Sansom, S. M. P.; Biggin, P. C. *Nature* **2001,** 414, 156.
2. Koga, K.; Gao, G. T.; Tanaka, H.; Zeng, X. C. *Nature* **2001**, 412, 802.
3. Koga, K.; Parra, R. D.; Tanaka, H.; Zeng, X. C. *J. Chem. Phys.* **2000**, 113, 5037.
4. Byl, O.; Liu, J-C.; Wang, Y.; Yim, W-L.; Johnson, J. K.; Yates, Jr, J. T. *J. Am. Chem. Soc.* **2006**, 128, 12090.
5. Maniwa, Y.; Kataura, H.; Abe, M.; Suzuki, S.; Achiba, Y.; Kira, H.; Matsuda, K. *J. Phys. Soc. Jpn.* **2002**, 71, 2863.
6. Maniwa, Y.; Kataura, H.; Abe, M.; Udaka, A.; Suzuki, S.; Achiba, Y.; Kira, H.; Matsuda, K.; Kadowaki, H.; Okabe, Y. *Chem. Phys. Lett.* **2005**, 401, 534.
7. Jackson, K. A.; Chalmers, B. *J. Appl. Phys.* **1958**, 29, 1178.
8. Maruyama, S. *Physica B* **2002**, 323, 193.
9. Maruyama, S, *Micro. Thermophys. Eng.* **2003**, 7, 41.
10. Shiomi, J.; Maruyama, S. *Phys. Rev. B* **2006**, 73, 205420.
11. Berendsen, H. J. C.; Grigera, J. R.; Straatsma, T. P. *J. Phys. Chem.* **1987**, 91, 6269.
12. Maruyama, S. *Handbook of Numerical Heat Transfer*; 2nd Ed. Wiley, 2006, 659.
13. Walther, J. H.; Jaffe, R.; Halicioglu, T.; Koumoutsakos, P. *J. Phys. Chem. B* **2001**, 105, 9980.
14. Brenner, D. W. *Phys. Rev. B* **1990**, 42, 9458.
15. Yamaguchi, Y.; Maruyama, S. *Chem. Phys. Lett.* **1998**, 286, 336.
16. Maruyama, S.; Taniguchi, Y.; Igarashi, Y.; Shiomi, J. *J. Therm. Sci. Tech.* **2006**, 1, 138.
17. Striolo, A.; Chialvo, A. A.; Gubbins, K. E.; Cummings P. T. *J. Chem. Phys.* **2005**, 122, 234712.
18. Kolesnikov, A. I.; Zanotti, J-M; Loong, C-K; Thiyagarajan, P. *Phys. Rev. Lett.* **2004**, 93, 035503.